\documentstyle[iopconf1,epsf]{article}

\def \be {\begin{equation}}
\def \ee {\end{equation}}

\begin{document}
\title{GENIUS: the first real-time detector  \\
for solar pp-neutrinos?}

\author{Laura\ Baudis \footnote{E-mail: Laura.Baudis@mpi-hd.mpg.de}, \\
H.\ V.\ Klapdor-Kleingrothaus \footnote{E-mail: 
          klapdor@daniel.mpi-hd.mpg.}}

\affil{
Max--Planck--Institut f\"ur Kernphysik, \\
P.O.Box 10 39 80\\ D--69029 Heidelberg, Germany}

\beginabstract
The GENIUS project is a proposal for a large supersensitive Germanium detector
system for WIMP and double beta decay searches with a much increased 
sensitivity relative to existing and other future experiments. 
In this paper, the possibility to detect low energy solar neutrinos
with GENIUS
in real-time through elastic neutrino-electron scattering is studied.
\endabstract

\section{Introduction}

The study of neutrinos coming from the Sun is a very active area of research.
Results from five solar neutrino experiments are now available.
These experiments measure the solar neutrino flux with different energy
thresholds and using very different detection techniques.
All of them, the Chlorine experiment at Homestake \cite{chlor},
the radiochemical Gallium experiments, GALLEX \cite{gallex} and SAGE 
\cite{sage}, the water Cerenkov detectors Kamiokande \cite{kamiok} and
Super-Kamiokande \cite{SuperK},
measure a deficit of the neutrino flux 
compared to the predictions of the standard solar models (SSM) \cite{SSM,dar}.
Recently it has been stated that it is impossible to construct
a solar model which would reconcile all the data \cite{hiroshi}. 
Moreover, a global analysis of the data of all the experiments does not leave
any room for the $^7$Be neutrinos \cite{Bah98b}.
On the other hand the  predictions of the SSM have 
been confirmed by helioseismology \cite{basu} to a high precision. 
An explanation of the results of solar neutrino experiments 
seems to require new physics beyond the standard model of electroweak 
interaction.

If neutrinos have non-zero masses and if they mix in analogy to the quark 
sector, then conversions between different neutrino flavours become
possible. Flavour conversions can occur in different physical scenarios,
depending on certain parameters on neutrino masses and mixing angles.
One oscillation scenario makes use of the 
MSW-mechanism \cite{msw85}, where the solar $\nu_e$ transform into other
neutrino flavours or into sterile neutrinos  as they pass through a thin  
resonance region near the solar core. 
The other scenario assumes that the neutrinos
oscillate in the vacuum between the Sun and the Earth \cite{gla87}, which means
that the oscillation length `just so` matches the Earth-Sun distance.

\section{The GENIUS Project}

GENIUS (GErmanium in liquid NItrogen Underground Setup) is a proposal
for operating a large amount of 'naked' Ge detectors in liquid nitrogen 
for dark matter and alternatively $\beta\beta$--decay researches 
\cite{ringb,hirsch,hellmig97,klap,Bau98,prop}, with an improved sensitivity 
of three orders of magnitude relative to present experiments. 

The proposed scale of the experiment 
is a nitrogen tank of about 12 m diameter and 12 m height 
which contains 100 kg (40 detectors) of natural Ge detectors
in its dark matter version, and 1000 kg of enriched $^{76}$Ge
detectors in a second step, for  neutrinoless double beta decay
searches. The liquid nitrogen acts both as a cooling
medium for the Ge detectors and as a shielding
against external radioactivity.
The optimal locations of the experiment would be the Gran Sasso
or the WIPP underground laboratories.  

In several technical studies it was shown, that Ge detectors work
reliably when immersed directly in liquid nitrogen
\cite{hellmig97,Bau98,prop}. The obtained performances for the energy
threshold and energy resolution
were as good as for conventionally operated detectors 
(i.e. vacuum-tight Cu-cryostat system \cite{kno89}).

To cover large parts of the MSSM parameter space, relevant for the
detection of neutralinos as the dark matter candidate \cite{jkg96,bedny1}, 
a maximum background
level of 10$^{-2}$ counts/(kg y keV) in the energy region between
0--100 keV has  to be achieved. 
This means a very large further background reduction
in comparison to our recent best result (20.82 counts/(kg y keV))
\cite{prl} with an enriched detector of the Heidelberg--Moscow
experiment \cite{hdmo} and to all other running dark matter
experiments.

\section{The solar neutrino spectrum}

\begin{figure}[h!]
\epsfxsize=10cm
\epsfbox{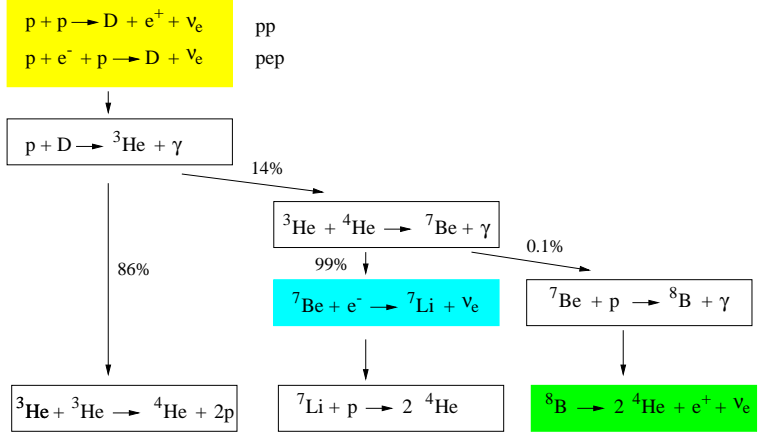}
\caption{Nuclear reactions in the pp-chain in the Sun.}
\label{ppchain}
\end{figure}

The Sun acquires its energy by nuclear reactions taking place in the core,
mainly via the so-called pp-chain (see Figure \ref{ppchain}).

\begin{figure}[h!]
\epsfxsize=12cm
\epsfbox{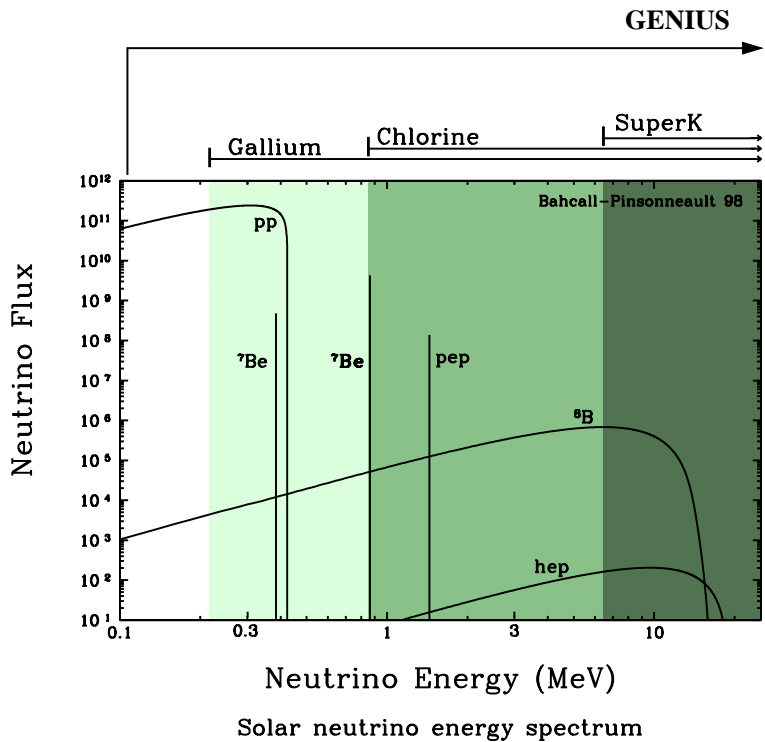}
\caption{Predicted solar neutrino spectrum in the SSM, from \cite{bah91}.}
\label{nuspec}
\end{figure}

The neutrino spectrum predicted by the SSM for the pp-chain
is shown in Figure \ref{nuspec}. The dominant part of the flux is
emitted at energies below 1~MeV.

The pp neutrinos, emitted in the reaction p+p $\rightarrow$ D+e$^+$+$\nu_e$,
have a continuous energy spectrum with the endpoint at 420 keV.
Their flux is most accurately predicted in the SSM, since it is strongly
restricted by the solar luminosity and by helioseismological measurements.
The other main features of the solar neutrino spectrum are a strong
monoenergetic line at 861 keV, from the reaction  $^7$Be+ e$^-$$\rightarrow$
$^7$Li+$\gamma$+$\nu_e$, the $^7$Be neutrinos, and a continuous
spectrum of neutrinos extending up to 15~MeV, due to the reaction
$^8$B$\rightarrow$2$\alpha$+e$^+$+$\nu_e$, the $^8$B neutrinos.
Table \ref{nufluxes} gives the solar neutrino fluxes in the SSM with
their respective uncertainties (from \cite{Bah98c}).

\section{Present status of the solar neutrino experiments}

\begin{figure}[h!]
\epsfysize=10cm
\epsfbox{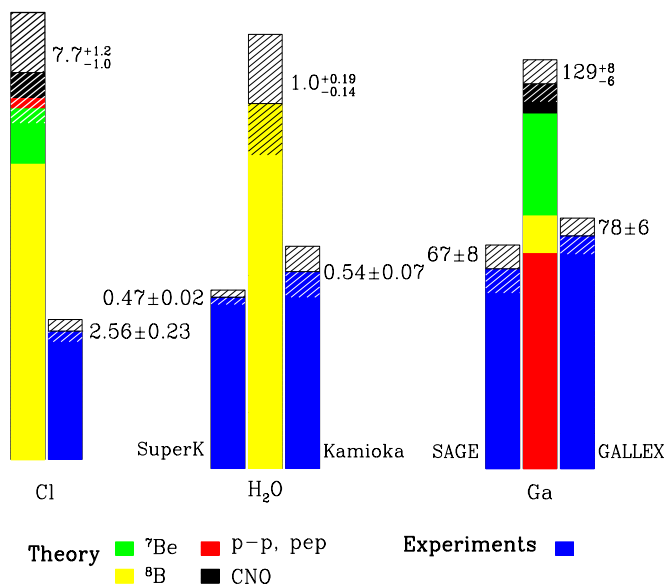}
\caption{Comparison of the total rates predicted in the SSM and the
observed rates in the present solar neutrino experiments, 
from \cite{Bahup}.}
\label{theoexp}
\end{figure}

The solar neutrino problem has been known for more than two decades,
since the Homestake experiment reported its first result.
At that time, however, it was not clear if the difference between
the chlorine measurement and the standard solar model prediction
was due to experimental systematics or the uncertainties in the 
SSM or if it was a sign of new physics.
Meanwhile, the observed discrepancy was confirmed by other
four solar neutrino experiments (see Figure \ref{theoexp}, from \cite{Bah96}).
Model independent analysis performed by many authors (see \cite{hiroshi}
and references therein) 
suggest that the solar neutrino problem can only be solved if
some additional assumptions are made in the standard electroweak
theory. The most generic assumption is to give neutrinos a mass,
which leads to neutrino oscillations in vacuum or matter.

Oscillations between two neutrino species are characterized by two
parameters: $\Delta$m$^2$, the difference of the squared mass eigenstates,
and $\theta$, the mixing angle between the mass eigenstates.
 
\begin{figure}[h!]
\epsfysize=10cm
\epsfbox{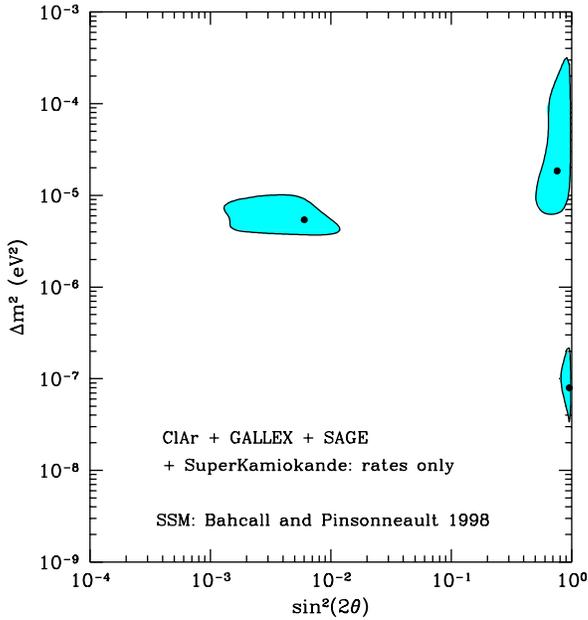}
\caption{The allowed  regions (99\% C.L.) in $\Delta m^2$ ---
$\sin^22\theta$ parameter space for the MSW solution, from \cite{Bah98}. }
\label{solutions}
\end{figure}

The Ga experiments, sensitive to the low-energy pp and $^7$Be neutrinos,
combined with the Homestake and Super-Kamiokande experiments, which
are sensitive to the high-energy $^8$B neutrinos, strongly restrict
the allowed range of $\Delta$m$^2$ and $\theta$ for all oscillation 
scenarios.
There exist four parameter areas compatible with the results of
all existing solar neutrino experiments: the  
large mixing angle solution (LMA), the small mixing angle solution (SMA), 
the low mass solution (LOW) and the vacuum oscillation solution
with strong mixing (see Figure \ref{solutions} for the MSW-solutions).
Up to date, there is no clear evidence for one of the above solutions.
To clarify the situation, there is great demand for additional solar
neutrino experiments, especially at energies below 1~MeV.

Borexino \cite{borexprop} is now being built up especially to measure
the flux of $^7$Be neutrinos in real-time. 
It will use 300 tons of organic scintillator
(100 tons of fiducial volume) to detect recoil electrons from
elastic neutrino-electron scattering. Since the scintillator has
no directional information and the signal is characterized only by the
scintillation light produced by the recoil electron, very stringent
constraints on the radiopurity of the scintilator and on the activity
of all detector materials are imposed.

\begin{table}
\hspace*{2.4cm}
\begin{tabular}{lc}
Source & Flux (10$^{10}$ cm$^{-2}$s$^{-1}$)\\
\hline
pp & 5.94 $\pm$ 0.01 \\
pep & 1.39$\times$10$^{-2}$$\pm$ 0.01 \\
$^7$Be & 4.80$\times$10$^{-1}$$\pm$ 0.09 \\
$^8$B & 5.15$\times$10$^{-4}$$\pm$ 0.19 \\
\end{tabular}
\caption{Solar Standard Model predictions of the neutrino fluxes, from
\cite{Bah98c}}
\label{nufluxes}
\end{table}

So far, there exist three proposals to measure the pp-flux in real-time, 
HERON \cite{heron}, HELLAZ \cite{hellaz} and LENS \cite{lens}.

The HERON project will use $^4$He in its superfluid state (at 20 mK) as the 
target medium. The detection reaction is elastic neutrino-electron
scattering, the electron recoil energy is converted into low-energy
elementary excitations of the helium, rotons, which can be detected.
For a fiducial volume of seven tons, the total SSM predicted event rate
is 14 per day (8 events per day from the pp neutrinos). HERON would
measure only the energy distribution of recoiling electrons, without
a direct determination of the neutrino energy.

In the HELLAZ project a large TPC (2000 m$^3$) filled with gaseous helium
at high pressure (5 atm.) and low temperature (77 K) will serve as a target.
It is planned to measure both the kinetic energy and the scattering angle
of recoil electrons from elastic neutrino-electron scattering and thus
to determine the solar neutrino energy.
The kinetic energy of recoil electrons is measured by counting the individual
electrons in a ionisation cloud generated by the energy loss of the recoil
electron due to ionisation in the helium gas.
The expected event rate for 2$\times$10$^{30}$ target electrons is 7 per day 
and 4 per day for pp neutrinos and $^7$Be neutrinos, respectively.

LENSE would be a complementary approach to the above detectors using flavour
independent elastic scattering from electrons.
The method of neutrino detection is neutrino capture in $^{82}$Se, 
$^{160}$Gd or $^{176}$Yb.
The neutrino captures occur to excited states of the final nuclides,
providing a strong signature against radioactive background.
The thresholds for neutrino capture are 173 keV for $^{82}$Se, 244 keV
for $^{160}$Gd and 301 keV for  $^{176}$Yb. Three different techniques
for implementation as a solar neutrino detector are explored at present
\cite{lenseloi}:
liquid scintillator loaded with Yb or Gd, scintillating crystals of silicates
of Gd (GSO) and time projection chambers with a gaseous compound of isotopic 
$^{82}$Se.

All of these projects are still in a stage of research and development,
they have not yet shown full feasibility for implementation as a solar
neutrino detector.

\section{ Time signatures of solar neutrinos}

Due to the eccentricity of the Earth orbit, seasonal variations
in the flux of solar neutrinos are expected.
The number of neutrinos of all flavours reaching the Earth is larger
when the Earth is closer to the Sun than when it is farther away and
should vary with 1/R$^2$, where R is the Earth-Sun distance,
R=R$_{0}$(1-$\epsilon$cos(2$\pi$t/year)). R$_{0}$= 1AU and $\epsilon$=0.017.
The neutrino flux thus shows a seasonal variation of about 7\%  from
maximum to minimum.
This variation can in principle be used by a real-time solar neutrino
experiment to extract the neutrino signal independently of background
(if the background is stable in time) and is limited only by 
statistics.

Beyond the so-called `normal' seasonal variation, an anomalous 
seasonal variation is predicted for the $^7$Be neutrino flux
in case of the vacuum oscillation solution,
since their oscillation length in this case is comparable to the seasonal
variation of the Earth-Sun distance due to the eccentricity of the 
Earth orbit.
The flux variations in this case are much larger than for the normal 
seasonal variation, 
they could serve as a unique
signature of vacuum oscillations \cite{gla87}.

\begin{figure}[h!]
\epsfysize=10cm
\epsfbox{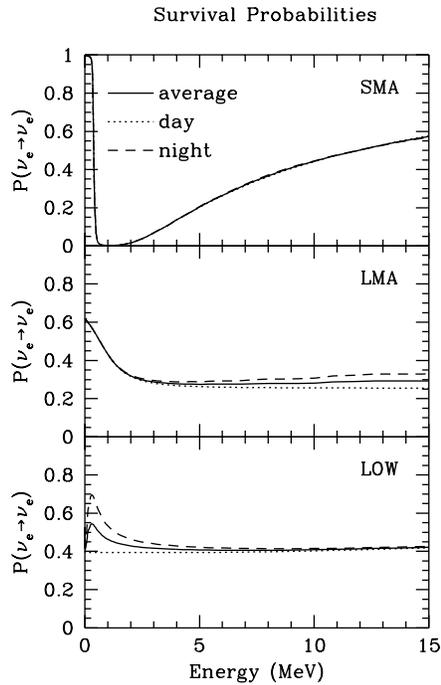}
\caption{Survival probabilities for an electron neutrino created in the Sun 
for the three MSW solutions, from \cite{bahkra}. SMA, LMA, LOW stand
for the small mixing angle, the large mixing angle and the low
$\Delta$m$^2$ MSW-solutions. }
\label{daynight}
\end{figure}

If neutrinos oscillate via the MSW-effect, then a regeneration of 
electron-neutrinos while passing through the Earth is predicted 
\cite{bahc89}. 
The so-called 'day/night'-effect is neutrino-energy dependent, its detection
would be a strong evidence for the MSW-effect. In Figure \ref{daynight}
(from \cite{bahkra}) the $\nu_e$ survival probabilities for the MSW solutions
computed for the day-time and night-time are shown. At
low energies only  the LOW solution shows significant differences between the
day- and night-time survival probability.
Therefore this solution could be tested by a real-time detector
of low energy solar neutrinos, in particular by measuring the pp and 
$^7$Be neutrino flux.

\section{GENIUS as a solar neutrino detector}

The goal of the GENIUS project as a dark matter detector is to
achieve the background level of 10$^{-2}$ events/kg y keV in the energy
region below 100~keV.  Such a low background in combination with
a target mass of at least  1 ton of natural (or enriched) Ge opens
the possibility to measure the solar pp- and $^7$Be-neutrino flux 
in real-time with a very low energy threshold.

\subsection{Signal Detection}

The detection reaction is  the elastic scattering process 
$\nu$ +  e$^- \rightarrow$ $\nu$ +  e$^-$.
The maximum electron recoil energy is 261 keV for the pp-neutrinos and 665 
keV for the  $^7$Be-neutrinos \cite{bahc89}. 
The energy of the recoiling electrons is detected through ionisation
in  high purity Ge detectors.
GENIUS in its 1~ton version would consist of an array of about 400 HPGe 
detectors, 2.5~kg each. Thus, the sensitive volume would be naturally divided 
into 400 cells which helps in background discrimination, since a 
neutrino interaction is taking place in a single cell.

\subsection{Signal Rates}

The dominant part of the signal in GENIUS is produced by 
pp-neutrinos (66 \%) and the  $^7$Be-neutrinos (33\%).

A target mass of 1 ton (10 tons) of natural or enriched Ge corresponds
to about 3$\times$10$^{29}$ (3$\times$10$^{30}$) electrons.

With the cross section for elastic neutrino-electron scattering 
\cite{bahc89}:\\

\noindent
$\sigma_{\nu_{e}}$ = 11.6 $\times$ 10$^{-46}$cm$^2$ \hspace*{3mm} pp\\
$\sigma_{\nu_{e}}$ = 59.3 $\times$ 10$^{-46}$cm$^2$ \hspace*{3mm} $^7$Be\\

\noindent
and the neutrino fluxes \cite{Bah98c}:\\

\noindent
$\phi_{pp}$ = 5.94 $\times$10$^{10}$ cm$^{-2}$s$^{-1}$\\
$\phi_{^{7}Be}$ = 0.48 $\times$10$^{10}$ cm$^{-2}$s$^{-1}$\\

\noindent
the expected number of events calculated  
in the standard solar model (BP98 \cite{SSM}) can be estimated:\\

\noindent
R$_{pp}$ = 68.9 SNU = 1.8 events/day (18 events/day for 10 tons)\\
R$_{^7Be}$ = 28.5 SNU = 0.6 events/day (6 events/day for 10 tons),\\

\noindent
The event rates for full $\nu_e \rightarrow \nu_{\mu}$ conversion 
are 0.48 events/day for pp-neutrinos and 0.14 events/day for
$^7$Be-neutrinos for 1 ton of Ge and ten times higher for 10 tons 
(see also Table \ref{rates})

\begin{table}[h!]
\begin{tabular}{lcc}
Case & Ev./day (11-665 keV) & Ev./day (11-665 keV)\\
&  (1 ton) & (10 tons)\\
\hline
SSM & 2.4  & 24 \\
Full $\nu_e \rightarrow \nu_{\mu}$ conv. & 0.62 & 6.2\\
\end{tabular}
\caption{Neutrino signal rates in GENIUS for 1 ton (10 tons) of 
Germanium.}
\label{rates}
\end{table}

\subsection{Background requirements}

GENIUS is conceived such that the external background from the natural 
radioactivity of the environment and from muon interactions is reduced
to a minimum, the main background contributions coming from the
liquid nitrogen shielding and the Ge detectors themselves.
To estimate the expected background counting rate, detailed Monte Carlo
simulations and calculations of all the relevant background sources 
were performed.
The sources of background can be divided into external and internal ones.
External background is generated by events originating from outside
the liquid shielding, such as photons and neutrons from the Gran Sasso
rock, muon interactions and muon induced activities.
Internal background arises from residual impurities in the liquid
nitrogen, in the steel vessel, in the crystal holder system, in the Ge crystals
themselves and from activation of both liquid nitrogen and Ge crystals
at the Earths surface.

For the simulation of muon showers, the external photon flux
and the radioactive decay chains the GEANT3.21 package 
\cite{geant} extended for nuclear decays was used.
This version had already successfully been tested in establishing a
quantitative background model for the Heidelberg--Moscow experiment
\cite{hdmo}. 

The results of the simulations are given in Table \ref{sim_solar} 
as counting rates per kilogramm detector material, year and keV 
for each simulated component together with the underlying assumptions 
for the background sources.

\begin{table}
\begin{tabular}{lllr}
Source & Component & Assumption & Events/\\
       &           &            & kg y keV     \\
       &           &            & 11-260keV \\
\hline
LiN, intrinsic  & $^{238}$U, $^{232}$T, $^{40}$K & 3.5, 4.4,
10$\times$10$^{-16}$g/g & 3.6$\times$10$^{-4}$\\ 
contamination & $^{222}$Rn  & 0.5 $\mu$Bq/m$^3$ &  2.5$\times$10$^{-5}$ \\        
\hline
Steel vessel   & U/Th & 10$^{-8}$g/g  &  4.5$\times$10$^{-5}$ \\
\hline
Holder system  & U/Th & 10$^{-13}$g/g; 13g/det. &8$\times$10$^{-5}$  \\
\hline
Surrounding    & Gammas      & GS flux; tank: 13$\times$13m &9$\times$10$^{-4}$ \\
               & Neutrons    &  GS flux & 3$\times$10$^{-4}$\\
               & Muon showers &  GS flux; muon veto 96\% &
               7.2$\times$10$^{-6}$ \\
               & $\mu$ $\rightarrow$ n ($^{71}$Ge)& 230
               capt. in nat. Ge/y  &  5$\times$10$^{-4}$  \\
               \hline
Cosmogens     &$^{54}$Mn,$^{57}$Co,$^{60}$Co
 & 1d activ., 5y deactiv.  & 8$\times10^{-4}$ \\
&$^{63}$Ni,$^{65}$Zn,$^{68}$Ge&\\
\hline
Total          & &  &3$\times$10$^{-3}$  \\
\end{tabular}
\caption{Simulated background sources together with the made
  assumptions and the resulting event rates in the low energy region of 
  the GENIUS project.}
\label{sim_solar}
\end{table}

In order for GENIUS to be sensitive to the low-energy solar neutrino
flux, a nitrogen shielding of 13 m in diameter is required.
Regarding the radiopurity of liquid nitrogen, the values reached at
present by the Borexino collaboration for their liquid scintillator
would be sufficient. Much attention has to be paid to the cosmogenic
activation of the Ge crystals at the Earth surface. In case of one day 
exposure, five years of deactivation below ground are required.
The optimal solution would be to produce the detectors in an underground 
facility.

\begin{figure}  
\centering 
\leavevmode\epsfxsize=300pt  
\epsfbox{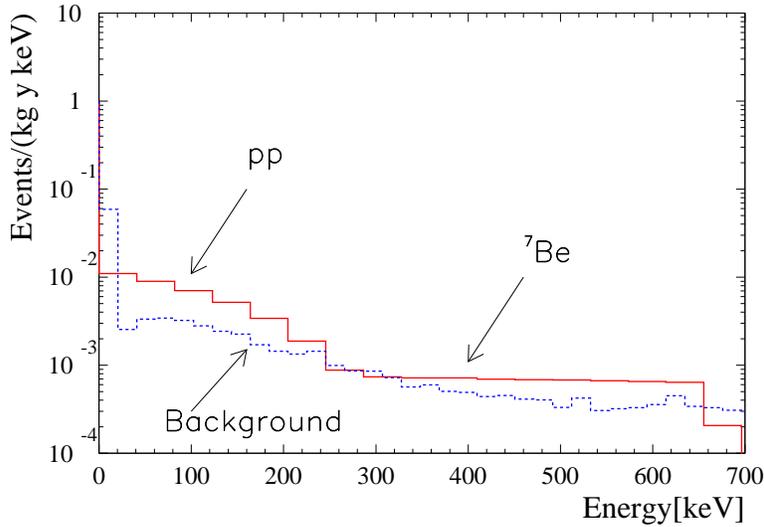}
\caption{\label{simspektrum} Simulated spectra of the low-energy 
neutrino signal (in the
SSM) and the total background in GENIUS (1 ton of natural germanium).}
\end{figure}

Figure \ref{simspektrum} shows the simulated spectrum of the low-energy neutrino
signal in GENIUS, together with the total expected background.

If the signal to background ratio S/B will be greater than 1, than the 
pp- and $^7$Be-neutrino flux can be measured by spectroscopic techniques
alone. 
If S/B $<$ 1,  one can make use of a solar signature in order to derive
the flux.

The eccentricity of the Earth's orbit induces a seasonal variation
of about 7\% from maximum to minimum. Even if the number of background
events is not known, the background event rate and the signal event rate
can be extracted independently by fitting the event rate to the seasonal
variation. The only assumption is that the background is stable in time
and that enough statistics is available. 

In case of a day/night - variation of the solar neutrino flux,
GENIUS would be sensitive to the LOW MSW solution of the
solar neutrino problem (compare Figure \ref{daynight}).

\section{Conclusion and Outlook}

GENIUS could be the first detector to detect the solar pp neutrinos
in real-time.
 Although this imposes
very strong purity restrictions for all the detector components, with a
liquid nitrogen shielding of 13 m in diameter and production of the Germanium 
detectors below ground, it should be feasible to achieve such a low
background level.
The advantages are the well understood detection technique
(ionization in a HPGe detector), the excellent energy resolution (1 keV 
at 300 keV), low energy threshold (about 11 keV) and the measurement
of the recoiling electrons in real-time.

The good energy resolution for detecting the recoiling electrons
would allow for the first time to measure the 1.3 keV predicted 
shift of the average energy of the beryllium neutrino line.
This shift is a direct measure of the central temperature of the Sun
\cite{bah93}.

\end{document}